\documentclass[a4wide,preprint,12pt,amsmath,amssymb]{elsarticle}

\usepackage{lineno,hyperref}
\usepackage[export]{adjustbox}

\journal{Journal of Physics and Chemistry of Solids}

\begin{document}
\newcommand{\etal}{{\sl et al.}}
\newcommand{\ie}{{\sl i.e.}}
\newcommand{\lto}{LaTiO$_3$}
\newcommand{\lao}{LaAlO$_3$}
\newcommand{\lno}{LaNiO$_3$}
\newcommand{\nith}{Ni$^{3+}$}
\newcommand{\nitw}{Ni$^{2+}$}
\newcommand{\otw}{O$^{2-}$}
\newcommand{\alo}{Al$_2$O$_3$}
\newcommand{\srtco}{SrTcO$_3$}
\newcommand{\aalo}{$\alpha$-Al$_2$O$_3$}
\newcommand{\xto}{$X_2$O$_3$}
\newcommand{\eg}{$e_{g}$}
\newcommand{\tg}{$t_{2g}$}
\newcommand{\dzt}{$d_{z^2}$}
\newcommand{\dxtyt}{$d_{x^2-y^2}$}
\newcommand{\dxy}{$d_{xy}$}
\newcommand{\dxz}{$d_{xz}$}
\newcommand{\dyz}{$d_{yz}$}
\newcommand{\egp}{$e_{g}'$}
\newcommand{\egone}{$e_{g}^1$}
\newcommand{\ag}{$a_{1g}$}
\newcommand{\mub}{$\mu_{\rm B}$}
\newcommand{\ef}{$E_{\rm F}$}
\newcommand{\alalo}{$a_{\rm Al_2O_3}$}
\newcommand{\asto}{$a_{\rm STO}$}
\newcommand{\nst}{$N_{\rm STO}$}
\newcommand{\lnnlam}{(LNO)$_N$/(LAO)$_M$}
\newcommand{\lxolao}{(La$X$O$_3$)$_2$/(LaAlO$_3$)$_4$(111)}
\newcommand{\xoalo}{($X_2$O$_3$)$_1$/(Al$_2$O$_3$)$_5$(0001)}

\begin{frontmatter}

\title{Interaction-driven spin-orbit effects and Chern insulating phases in  corundum-based $4d$ and $5d$ oxide honeycomb lattices}
\author[a]{Okan K\"oksal} 
\author[a]{Rossitza Pentcheva\corref{Rossitza}}
\address[a]{Department of Physics and Center for Nanointegration Duisburg-Essen (CENIDE), University of Duisburg-Essen, Lotharstr. 1, 47057 Duisburg, Germany}
\cortext[Rossitza]{Corresponding author}
\ead{Rossitza.Pentcheva@uni-due.de}

\begin{abstract}

Using density functional theory calculations with a Hubbard $U$, we explore topologically nontrivial phases in  $X_2$O$_3$ honeycomb layers with $X=$ $4d$ and $5d$ cation inserted in the band insulator  $\alpha$-Al$_2$O$_3$ along the [0001]-direction.  Several promising candidates for quantum anomalous Hall insulators (QAHI) are identified. In particular, for $X$\,=\,Tc and Pt spin-orbit coupling (SOC) opens a gap of 54 and 59 meV, respectively, leading to Chern insulators (CI) with $C$\,=\,--2 and --1. The nature of different Chern numbers is related to the corresponding spin textures. The Chern insulating phase is sensitive to the Coulomb repulsion strength: $X$\,=\,Tc undergoes a transition from a CI to a trivial metallic state beyond a critical strength of $U_c =2.5$ eV. A comparison between the isoelectronic metastable FM phases of $X$\,=\, Pd and Pt emphasizes the intricate balance between electronic correlations and SOC: while the former is a trivial insulator, the latter is a Chern insulator. In addition, $X$\,=\,Os turns out to be a FM Mott insulator with an unpaired electron in the \tg\ manifold where SOC induces an unusually high orbital moment of 0.34 \mub\ along the $z$-axis. Parallels to the $3d$ honeycomb corundum cases are discussed. 

\end{abstract}

\end{frontmatter}


\section{Introduction}
\label{S:1}

Chern insulators -- the time-reversal symmetry (TRS) broken analog of  topological insulators (TI) -- are characterized by a quantized anomalous Hall conductivity even in the absence of an external magnetic field \cite{Weng2015,Ren2016} and are potential candidates for application in low-power electronic devices and for the realization of Majorana fermions. 
Haldane's prediction of a quantum Hall effect in the absence of an external magnetic field for spinless fermions on a honeycomb lattice shifted the interest towards the latter \cite{Haldane}. Much effort has been directed in the past years to identify materials realizations, in particular such that host the honeycomb pattern. One strategy is to dope magnetic impurities into known topological insulators in order to break TRS, as e.g.  Mn-doped HgTe or Cr-, Fe-doped Bi$_{2}$Te$_{3}$, Bi$_{2}$Se$_{3}$, Sb$_{2}$Te$_{3}$ \cite{Liu_Zhang,Yu_Zhang,Fang_Bernevig} or $5d$ transition metals on graphene \cite{Zhang2012,Zhou_Liu} or OsCl$_3$ \cite{Sheng2017}.  On the other hand, an interesting class of materials to consider are transition metal oxides (TMO). Besides the tendency towards TRS breaking and larger band gaps than in conventional $sp$ bonded systems, their intricate interplay of spin, orbital and lattice degrees of freedom is encouraging to add further functionalities and richer behavior. CI have been sought e.g. among rocksalt- (EuO/CdO \cite{Zhang2014} and EuO/GdO \cite{Garrity2014}) or rutile-derived heterostructures \cite{Huang_Vanderbilt,Cai_Gong,Lado2016} as well as  pyrochlore structures \cite{Fiete2015}.  As proposed recently by Xiao et al. \cite{Xiao2011}, a buckled honeycomb pattern is formed in (111)-oriented A$X$O$_3$ perovskite superlattices (SL) by each pair of triangular $X$-layers. Several perovskite-derived candidates for TI were proposed as e.g. SrIrO$_3$ and LaAuO$_3$ bilayers, however the former shows a tendency towards an antiferromagnetic (AFM) coupling if correlation effects are considered \cite{Lado2013,Okamoto2014}. A systematic study of the $3d$ series in (La$X$O$_3$)$_2$/(LaAlO$_3$)$_4$(111)\cite{Doennig2016} identified a particularly strong effect of spin-orbit coupling for LaMnO$_3$ unanticipated in $3d$ compounds, showing the system is a Chern insulator with a significant gap of ~150 meV when the two sublattices of the honeycomb pattern are symmetric. However, the system turns out to be unstable with respect to a Jahn-Teller (JT) distortion, the ground state thus being a trivial Mott insulator. One strategy to stabilize the CI phase is to suppress the JT distortion e.g. by selective excitation of phonons, an approach used recently to induce an insulator-to-metal transition in NdNiO$_3$/LaAlO$_3$(001) SLs \cite{caviglia12}. Another strategy is to move from the $3d$ elements where correlation effects are notoriously strong, while SOC is weaker, to $4d$ and $5d$ elements, where the balance between correlations and SOC can even be reversed. In particular $4d$ and $5d$ systems show a much weaker tendency towards symmetry breaking transitions like the above mentioned JT effect, common to the strongly correlated $3d$ cases. This design strategy led to the identification of LaRuO$_3$ and LaOsO$_3$ honeycomb bilayers sandwiched in LaAlO$_3$(111) as robust QAHI \cite{HongliNQM}. It is important to note the low spin configuration of  Ru$^{3+}$ and Os$^{3+}$ with a single hole in the  \tg\ states, in contrast to the isoelectronic high spin Fe$^{3+}$ ($d^{5}$) in LaFeO$_3$ with AFM ground state \cite{HongliNQM}.

\begin{figure}[h!]
\centering
\includegraphics[width=8.8cm,keepaspectratio]{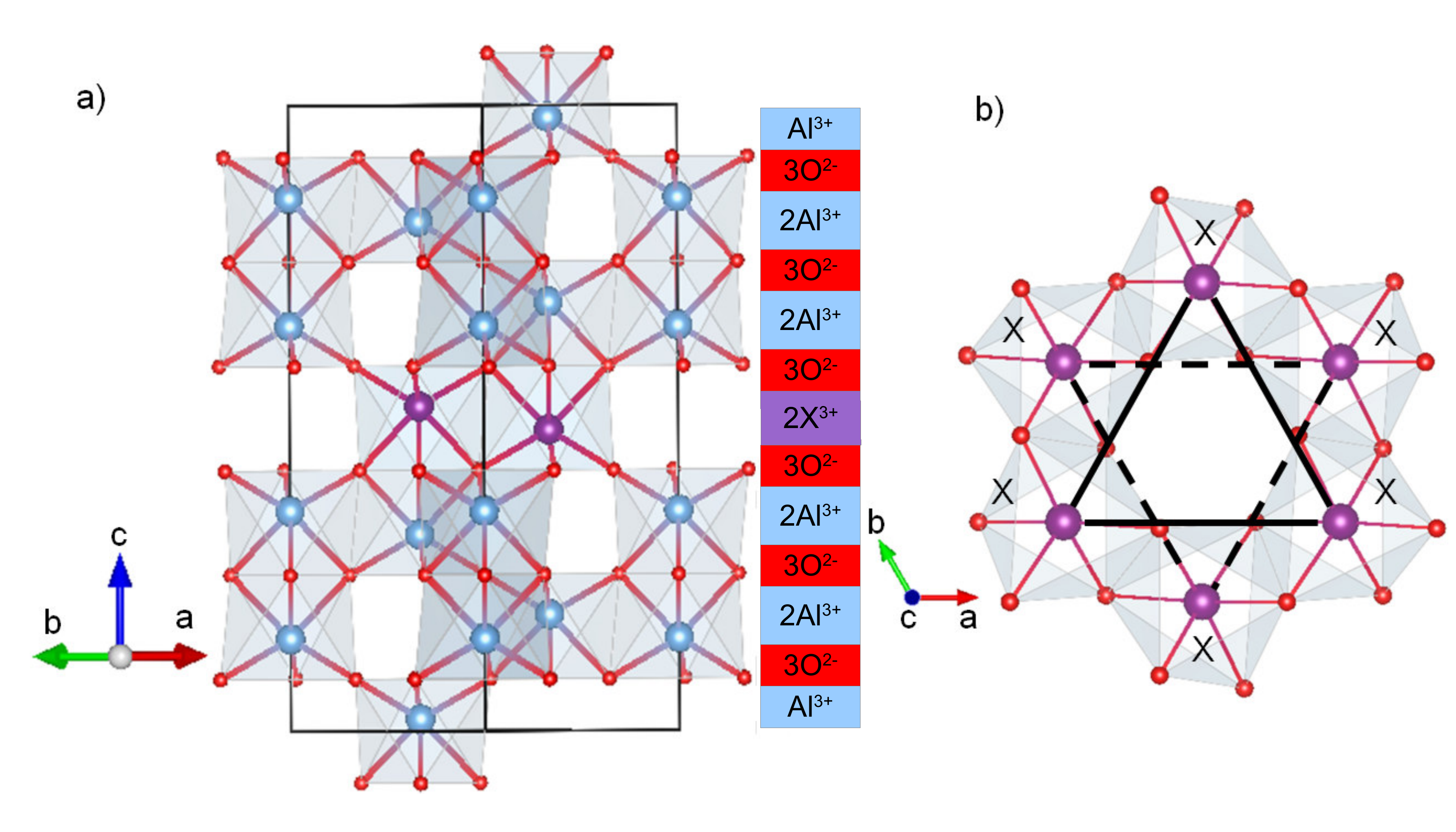}
\caption{a) Side view of the ($X$$_{2}$O$_{3}$)$_{1}$/(Al$_{2}$O$_{3}$)$_{5}$(0001) superlattice where $X=4d$ and $5d$ ion. b) Top view of the buckled honeycomb lattice in $a$-$b$ plane, solid and dashed lines connect the next nearest TM-ion neighbors residing on the two sublattices.} 
\label{fig:schematic_view}
\end{figure}

Another crystal structure that naturally hosts the honeycomb lattice is the corundum, \aalo. Here each cation monolayer forms a buckled honeycomb layer, though the degree of buckling is significantly reduced compared to the perovskite. Moreover, also the type of connectivity is different: while in the perovksite case only corner sharing is present, in the corundum the $X$O$_6$ octahedra in the $X_2$O$_3$ layer are edge-sharing and alternating  corner- and face-sharing to the next layer above and below (cf. Fig. \ref{fig:schematic_view}).  Our previous work has addressed the complex electronic behavior of  \xoalo(0001) with $X=3d$ \cite{OKRP2016} where a variety of electronic phases was identified that  strongly differs from the corresponding bulk $X_2$O$_3$ compounds.  While in most cases we find AFM ground states, the metastable ferromagnetic cases of $X$=Ti, Mn, Co and Ni with constrained symmetry of the two sublattices showed a common band structure of four bands, two flat and two with a Dirac-like crossing at $K$ close to \ef, associated with the single band filling of \egp$^1$ (Ti) or  \eg$^1$ (Mn, Ni and Co) states.\footnote{Note that the trigonal symmetry splits the \tg\ states in an \ag\ and doubly degenerate \egp\ states.} A similar band structure was also obtained for the analogous perovskite cases of $X$=Mn, Co and Ni \cite{Doennig2016}. Applying lateral strain it was possible to tune the Dirac point to the Fermi level for the corundum $X$=Mn, Co and Ni, however SOC showed a weak effect with a gap opening of several meV, but still with considerable nearly integer anomalous Hall conductivity, arising from the high Berry curvature around the $K$ points. On the other hand we obtained one case ($X$\,=\,Ti) of extremely strong SOC at the lateral lattice constant of \aalo\ within GGA ($a=4.81$ \AA). For $X=$ Ti SOC with magnetization along the [0001]-direction leads to an almost integer orbital moment ($-0.88$\mub), which is antialigned to and almost compensates the spin-moment of 1.01~\mub, pointing towards a likely realization of Haldane's model of spinless fermions on a honeycomb lattice. SOC acts here as a Zeeman term and lifts the degeneracy of the half-filled \egp\ orbital leading to a metal-to-insulator transition with a strong rearrangement of bands. A similar situation arises in BaFe$_2$(PO$_4$)$_2$ (BFPO) \cite{Warren2015,Kee2017}, where Fe$^{2+}$ has a filled majority spin band and a single electron with \egp\ orbital polarization in the minority band. Although (Ti$_2$O$_3$)$_1$/(Al$_2$O$_3$)$_5$(0001) has zero Chern number, two pairs of nontrivial bands with integer but opposite Chern numbers are identified above and below \ef. Suitable doping or electric field may eventually shift the Fermi level and turn the system into a quantum anomalous Hall insulator. This strong effect of SOC significantly reduces the energy difference to the stable AFM ground state by one order of magnitude. 

Although some interesting ground and metastable states were found for the $3d$ corundum-derived superlattices, the emergence of stable Chern insulating phases was not observed. This has motivated us to explore here the analogous $4d$ and $5d$ systems. Previously, Afonso and Pardo \cite{Pardo2015} studied $5d$ honeycomb layers sandwiched in the corundum structure, concentrating only on TI cases that preserve time reversal and inversion symmetry. The non-magnetic Au$_2$O$_3$ with $d^8$ configuration (analogous situation to the high spin $d^4$ case discussed above e.g. for $X=$ Mn but with single occupation of \eg\ states in each spin channel) was indentified as a potential candidate for $Z_2$ TI. Still, the latter topologically nontrivial configuration transforms into a trivial one with increasing $U$ or tensile strain. Moreover, it is metastable, the ground state is found to be AFM for realistic and even negative $U$ values ($U>-3$~eV) and  \cite{Pardo2015}. 

Here we are interested in particular in possible QAHI phases, therefore we lift the constraint of TRS  in $4d$ and $5d$ corundum-derived systems. Indeed we find several promising cases where a CI phase can be stabilized, e.g. for $X=$ Tc, Pt. For Tc$_2$O$_3$ we explore the dependence of QAHI on the Coulomb repulsion parameter, showing a transition from a CI to a trivial Mott insulator beyond a critical value $U_c =2.5$ eV. Moreover, we discuss the sensitive balance between electronic correlations and SOC in the case of the isoelectronic $X=$ Pd and Pt, the former being a trivial Mott insulator and the latter a CI. Last but not least, we address Os$_2$O$_3$, which shows a substantial orbital moment.

\section{Theoretical methods}

Density functional calculations were carried out for ($X_{2}$O$_{3}$)$_{1}$/(Al$_{2}$O$_{3}$)$_{5}$(0001), $X= 4d$ or $5d$ ion, with the VASP \cite{VASP} code using the projector augmented wave (PAW)  method\cite{PAW}. For the exchange-correlation functional we applied the generalized gradient approximation (GGA) of  Perdew, Burke and Enzerhof \cite{GGA_PBE}. An on-site Coulomb repulsion parameter was considered within the GGA\,+$U$ approach as implemented by Liechtenstein et al. \cite{Liechtenstein} with $J=0$ eV. The values of $U$ for $4d$ and $5d$ ions are generally much lower than for $3d$, in the range of $0-3$ eV. The dependence of the topological properties of $X=$ Tc on the value of $U$ is further addressed in Section \ref{sec_Tc_U_dep}. The hexagonal simulation cell, shown in Fig.  \ref{fig:schematic_view}, contains 30 atoms (18 oxygens, two $X$ and ten Al cations). The calculations were performed using a $k$-point mesh of 10$\times$10$\times$2 including the $\Gamma$-point. The optimized lattice constants of \aalo\ within GGA are $a$=\,4.81\,\AA,\,$c$\,=\,13.12 \AA\,being $\sim1$\% larger than the experimentally obtained values of $a$=\,4.76\,\AA\,and \,$c$\,=\,12.99 \AA \cite{Newnham1962}. A cutoff energy of 600 eV was used for the plane waves. Atomic positions were relaxed until the Hellman-Feynman forces were lower than 1 meV/\AA. In particular, high-symmetry metastable or symmetry-broken states were explored also with the all-electron full-potential linearized augmented plane wave (LAPW) method as implemented in the Wien2k code \cite{wien2k}. The anomalous Hall conductivity (AHC) for potentially interesting cases is computed on a very dense $k$-point mesh of 144$\times$144$\times$12 using the wannier90 code \cite{wannier90}.

\section{Results and discussion}

Here we extend our previous study of corundum-derived honeycomb lattices \cite{OKRP2016} to $4d$ and $5d$ cases and identify several candidates for Chern insulators. Using the insight obtained for the $3d$ series, the discussion is extended to the $4d$ and $5d$ homologous elements of Mn, Fe and Ni. For $X$= Rh and Ir only a non-magnetic solution,  leading to trivial insulators, is found within DFT+$U$. In contrast, for Re and Ru ferromagnetic solutions were obtained, but SOC is unable to open a gap. We concentrate instead on $X=$ Tc and Os where the ferromagnetic solutions are favoured by 116 meV and 18 meV per u.c. (unit cell) for $U=1.0$ eV, respectively, compared to the non-magnetic case. For $X=$ Pd and Pt a ferromagnetic phase can be stabilized, but is 0.28 eV and 0.67 eV per u.c. higher in energy than the respective antiferromagnetic and non-magnetic ground states. Thus, in the following we discuss the electronic properties of the FM phases of Tc, Pd, Pt and Os.
\begin{figure} [h!]
\centering
\includegraphics[width=14.0cm,keepaspectratio]{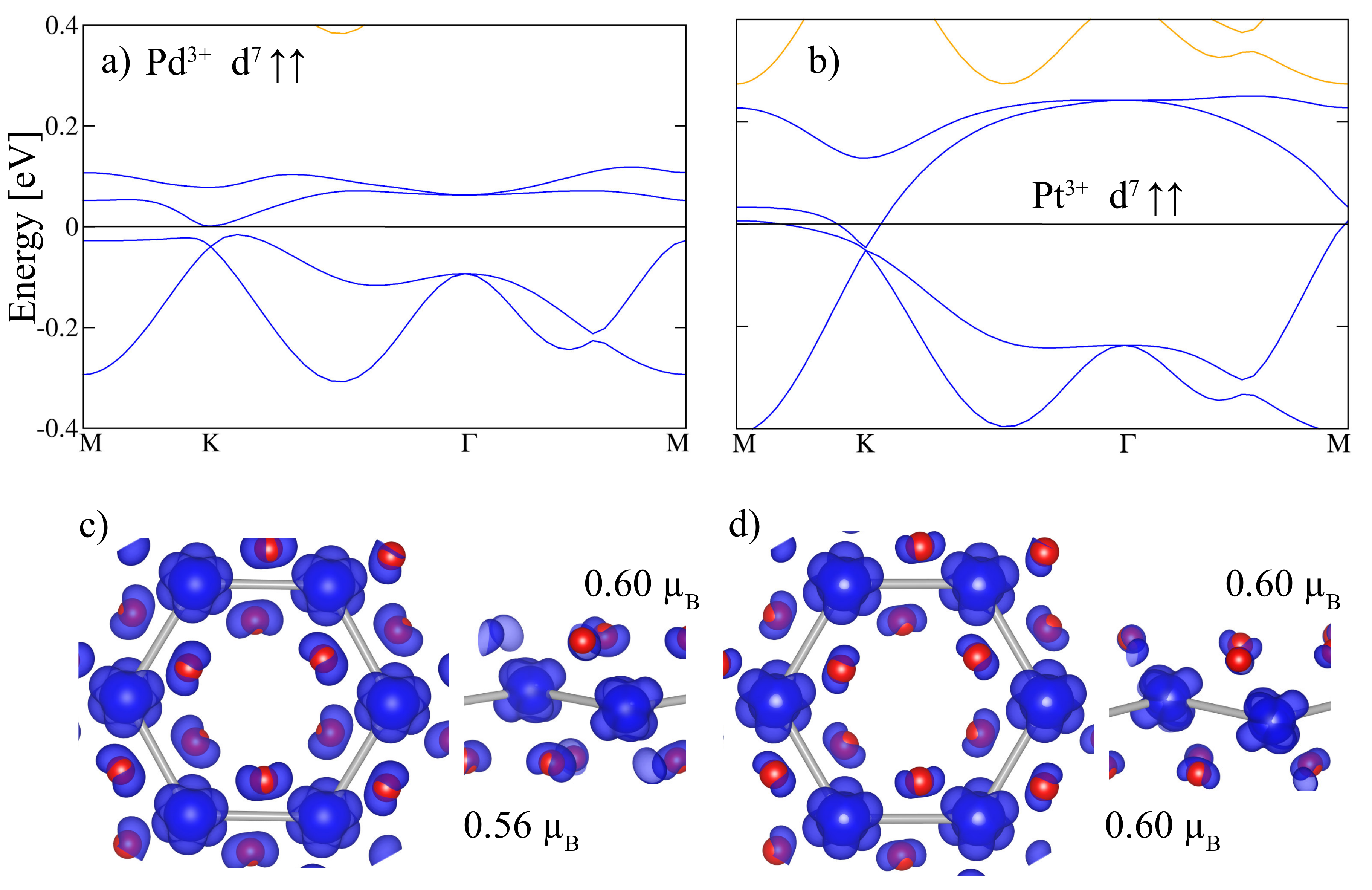}
\caption{Spin-resolved band structure of \xoalo(0001), with a) $X$\,=\,Pd ($U=1.0$ eV, $a= 4.81$\AA) and b) Pt ($U=0.5$ eV, $a= 4.76$ \AA). Blue/orange denote majority/minority bands and the Fermi level is set to zero. c) and d)  top and side view of isosurfaces of the spin density, integrated in the energy range of --1 eV to \ef,  blue (red) show the  majority (minority) contributions.}
\label{fig:Pd_Pt}   
\end{figure}

\subsection{GGA+$U$(+SOC) results for isoelectronic $Pd_{2}$O$_{3}$ and $Pt_{2}$O$_{3}$}


The band structure of the isolectronic $X$= Pd and Pt around \ef\ is dominated by four majority bands. While  $X$= Pd  is a semiconductor with an indirect band gap of $\sim17$ meV and a narrower bandwidth of $\sim0.4$ eV (cf. Fig. \ref{fig:Pd_Pt}a), $X$= Pt is metallic and the four characteristic majority bands show a larger bandwidth of $\sim0.6$ eV (cf. Fig. \ref{fig:Pd_Pt}b). Both the bandwidth and tendency towards metallic behavior is consistent with the larger extension of the $5d$ orbitals as compared to the $4d$. The spin-densities indicate substantial occupation of both \eg\ orbitals, signaling a $d^8L$ configuration, similar to the one of the Ni-ion in LaNiO$_3$ nickelates \cite{romero,Freeland:11,Doennig2014} rather than the formal $d^7$ occupation. 


\begin{figure} [ht!]
\centering
\includegraphics[width=12.0cm,keepaspectratio]{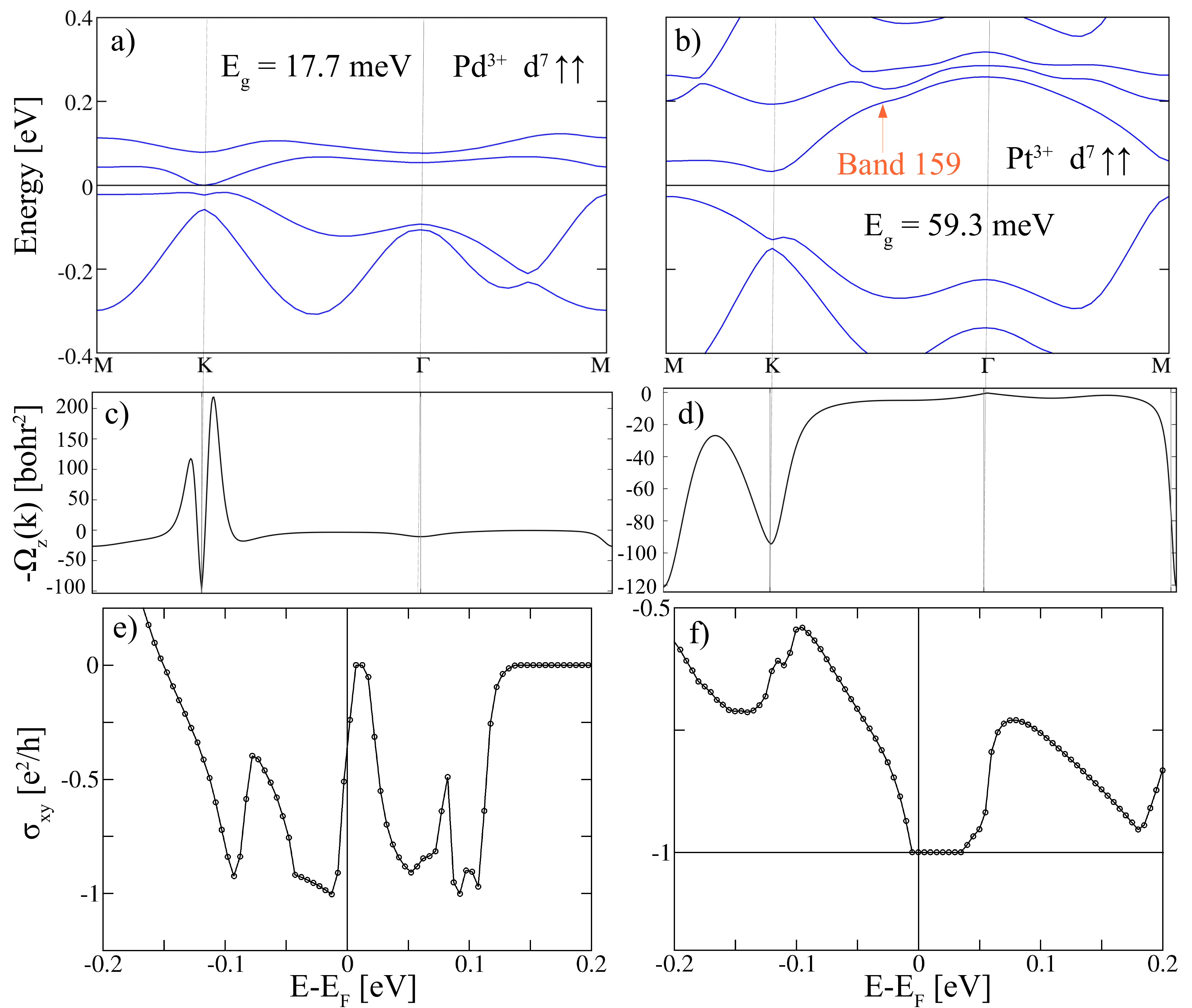}
\caption{ a) and b) GGA\,+\,\textit{U}\,+\,SOC band structures for the isovalent and isoelectronic \xoalo, $X$\,=\,Pd ($U=1.0$ eV) and \,Pt ($U=0.5$ eV) with magnetization along the [0001] direction as well as the Berry curvatures c) and d) along the same $k$-path. e) and f) show the corresponding anomalous Hall conductivity $\sigma_{xy}$ in units of $e^{2}/h$ as a function of the chemical potential.}
\label{fig:Pd_Pt_SOC_AHC}   
\end{figure}

\begin{figure} [htb!]
\centering                                                   
\includegraphics[width=7.0cm,keepaspectratio]{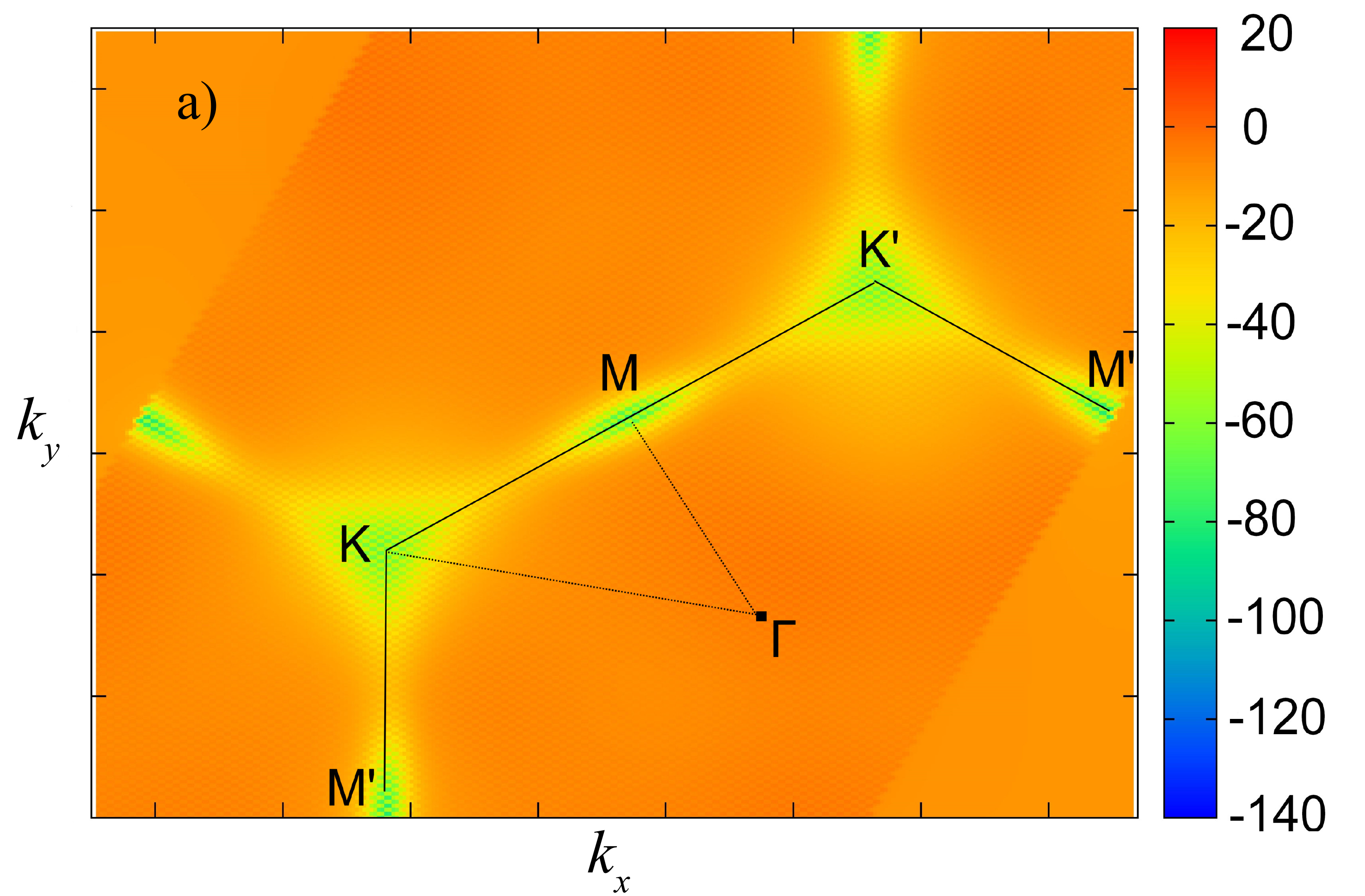}
\caption{a) Top view of the Berry curvature  $\Omega(k)$ for (Pt$_{2}$O$_{3}$)$_{1}$/(Al$_{2}$O$_{3}$)$_{5}$(0001) at $U$\,=\,0.5 eV in the Chern-insulating phase.}
\label{fig:BC_Pt}   
\end{figure}

We proceed with the effect of SOC on the latter two cases. 
Inclusion of SOC for the Pd honeycomb corundum monolayer does not alter the band gap (cf. Fig. \ref{fig:Pd_Pt_SOC_AHC}a) but leads to an avoided crossing of the two occupied bands at K and $\Gamma$ and a splitting of the empty pair of bands at $\Gamma$, resulting in strong oscillations of the Berry curvature at $K$  between positive and negative values (Fig. \ref{fig:Pd_Pt_SOC_AHC}c) that lead to a nonvanishing $\sigma_{xy}$ but with a low  value at \ef. In contrast, for Pt SOC opens a gap of 59 meV at M and lifts the degeneracy of bands both at K and $\Gamma$ (Fig. \ref{fig:Pd_Pt_SOC_AHC}b). The corresponding Berry  curvature $\Omega(k)$ shows strong negative peaks around M and K (Fig. \ref{fig:Pd_Pt_SOC_AHC}d), that are also clearly visible in the top view of $\Omega(k)$ in Fig. \ref{fig:BC_Pt}. A Chern insulating phase with $C$\,=\,--1  is found for the experimental lattice parameter of \alo, $a= 4.76$ \AA\ \cite{Newnham1962} with a broad Hall plateau at \ef\ (Fig. \ref{fig:Pd_Pt_SOC_AHC}f). For a slightly larger lateral lattice constant $a= 4.81$ \AA\ (the GGA value) the highest valence band crosses the Fermi level at M and the system becomes metallic  (not shown here).  For  comparison, a $Z_2$ TI phase was found in the nonmagnetic state of the perovskite-derived (KTaO$_3$)$_9$/(KPtO$_3$)$_2$(111) SL \cite{Lado2013}.  The latter is found to  be stable for realistic values of $U$, but the system becomes AFM beyond $U_c> 3$. In contrast to the corundum case here, no FM solution could be stabilized. 

\subsection{Tc$_2$O$_3$: transition from a Chern to a metallic phase with $U$}
\label{sec_Tc_U_dep}

\begin{figure} [h!]
\centering
\includegraphics[width=11.0cm,keepaspectratio]{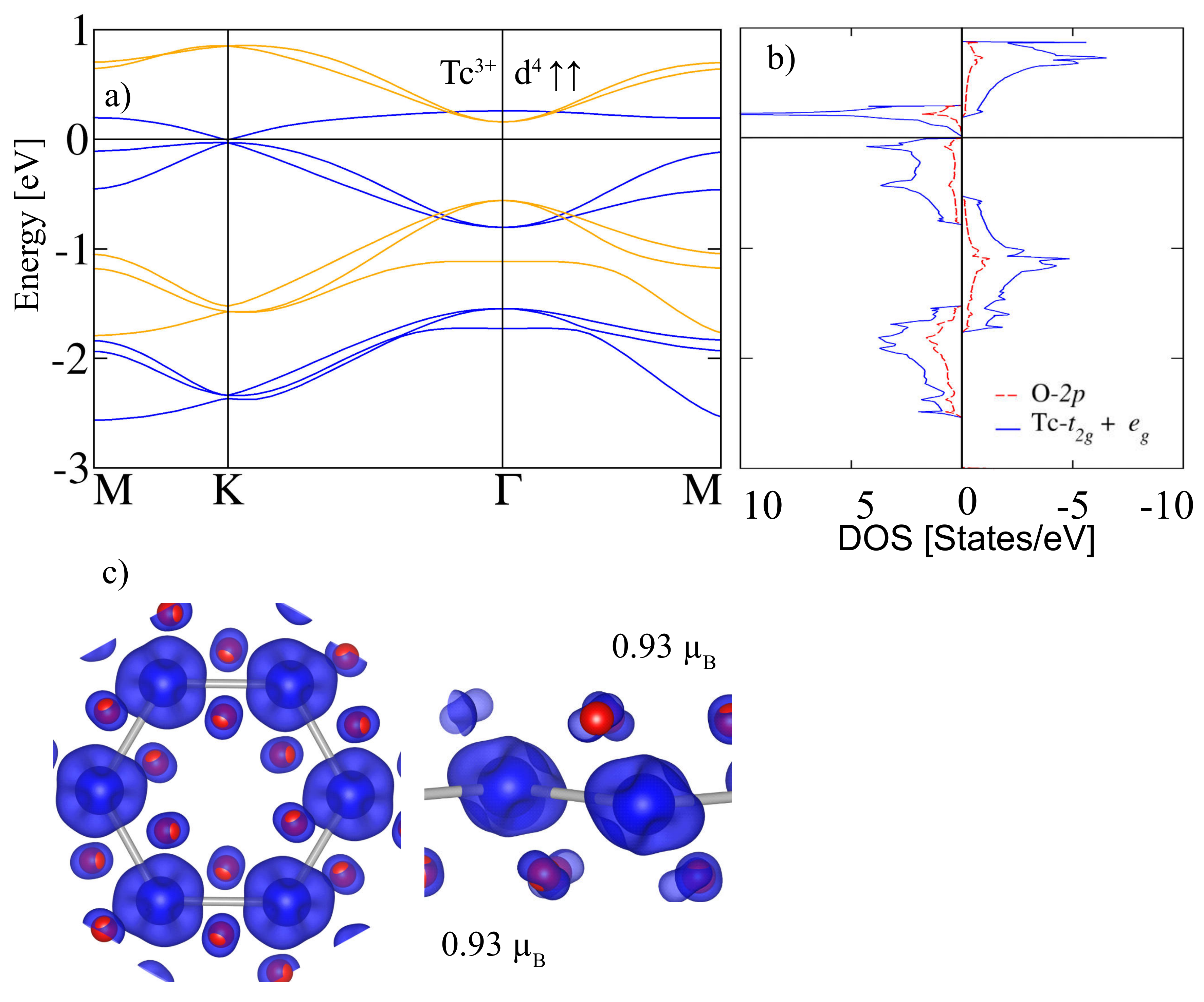}
\caption{a) Spin-resolved band structure and spin density distribution integrated in the energy range of --3 eV to \ef\ for \xoalo, $X$\,=\,Tc at a lateral lattice constant $a= 4.81$ \AA\ for $U=1.0$ eV.  Both the spin density in (c) and the  density of states in (b) display the strong hybridization between O $2p$ and Tc $4d$ states. Positive (negative) values of the  density of states correspond to the majority (minority) states. Same color coding is used in c, d) as in Fig. \ref{fig:Pd_Pt}.}
 
\label{fig:Tc_DOS}   
\end{figure}

As illustrated in Fig. \ref{fig:Tc_DOS}a, for $U=1.0$ eV without SOC (Tc$_{2}$O$_{3}$)$_{1}$/(Al$_{2}$O$_{3}$)$_{5}$(0001) is a semimetal with majority bands nearly touching at $K$, whereas a pair of bands, degenerate at $\Gamma$, comprises the conduction band minimum in the minority channel. In  the $4d^4$ configuration of Tc$^{3+}$ all electrons are in the \tg\ subshell. According to the Hund's rule a magnetic moment of 2 \mub\ is expected due to two unpaired electrons, but owing to the large bandwidth of the Tc \tg\ band ($\sim2.5$ eV, cf. Fig. \ref{fig:Tc_DOS}a) and the strong hybridization between Tc $4d$ and O $2p$, visible from the spin density in Fig. \ref{fig:Tc_DOS}c, the calculated value is  much lower  (0.93 \mub). Similar reduction of the magnetic moment has been reported also in other Tc-compounds, as e.g. $A$TcO$_3$ perovskites ($A$=Sr, Ba, Ca) both experimentally \cite{Maxim2011,Rodriguez2011} and theoretically  \cite{Franchini2011,Georges2012}, as well as for  $A$TiO$_3$/$A$TcO$_3$(001) superlattices \cite{Vladislav2015}. 

\begin{figure} [hp!]
\includegraphics[width=13.5cm,keepaspectratio]{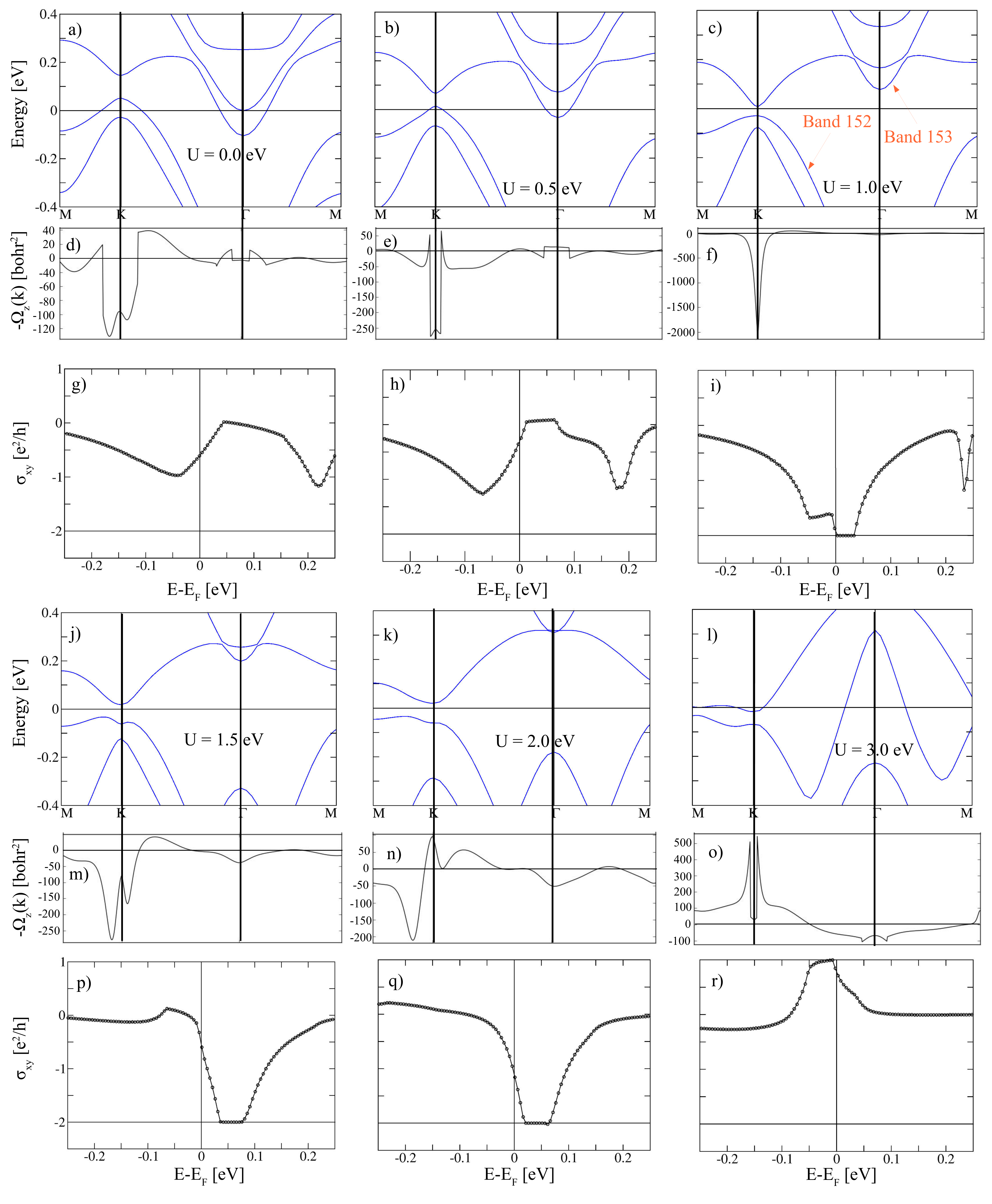}
\caption{GGA\,+\,\textit{U}\,+\,SOC results for (Tc$_{2}$O$_{3}$)$_{1}$/(Al$_{2}$O$_{3}$)$_{5}$(0001) as a function of the on-site Coulomb repulsion parameter $U$:  Evolution of the band structure (a-c, j-l), Berry curvatures (d-f, m-o) plotted along  the same $\textit{k}$-path  and AHC $\sigma_{xy}$ vs. the chemical potential (g-i and p-r).}
\label{fig:Tc_BC_AHC}   
\end{figure}

In the following, we analyze the topological properties of (Tc$_{2}$O$_{3}$)$_{1}$/(Al$_{2}$O$_{3}$)$_{5}$(0001) as a function of $U$. Fig. \ref{fig:Tc_BC_AHC} displays the band structures, Berry curvatures $\Omega(k)$ plotted along the same $k$-path and the anomalous Hall conductivity versus the chemical potential as afunction of the Hubbard repulsion parameter $U$. 
  %
%
%
\begin{figure} [htb!]
\centering                                                   
\includegraphics[width=13.0cm,keepaspectratio]{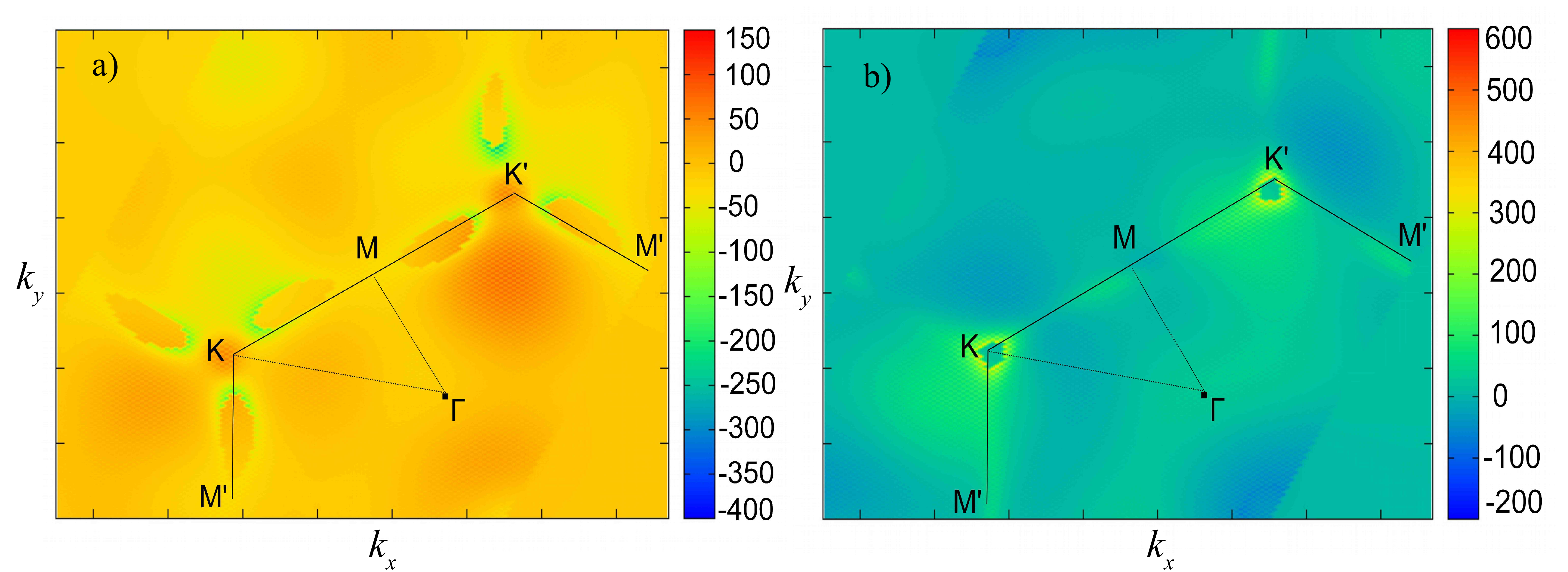}
\caption{Top views of the Berry curvature distribution $\Omega(k)$ for (Tc$_{2}$O$_{3}$)$_{1}$/(Al$_{2}$O$_{3}$)$_{5}$(0001) at $U$\,=\,2.0 eV in the Chern-insulating phase and trivial phase at $U$\,=\,3.0 eV displayed in a) and b).}
\label{fig:BC_Tc}   
\end{figure}
In the GGA case, the main contribution to $\Omega(k)$ (Fig. \ref{fig:Tc_BC_AHC}d) results from the bands around K: the valence band overlap with the Fermi level at K gives rise to a splitting into two peaks in the corresponding Berry curvature. At $U=0.5$ eV these two peaks around K sharpen (cf. Fig. \ref{fig:Tc_BC_AHC}e). The conduction band at $\Gamma$ crosses \ef\ and forms an electron pocket (Fig. \ref{fig:Tc_BC_AHC}a, b). With increasing $U$ this band (153) is shifted to higher energies and a gap opens at $U=1.0$ eV. Moreover, the two occupied bands crossing at K split, pointing to an enhanced effect of SOC due to electronic correlation (cf. Fig. \ref{fig:Tc_BC_AHC}k for $U=2.0$ eV). 

\begin{figure} [ht!]
\includegraphics[angle=0,scale=0.20]{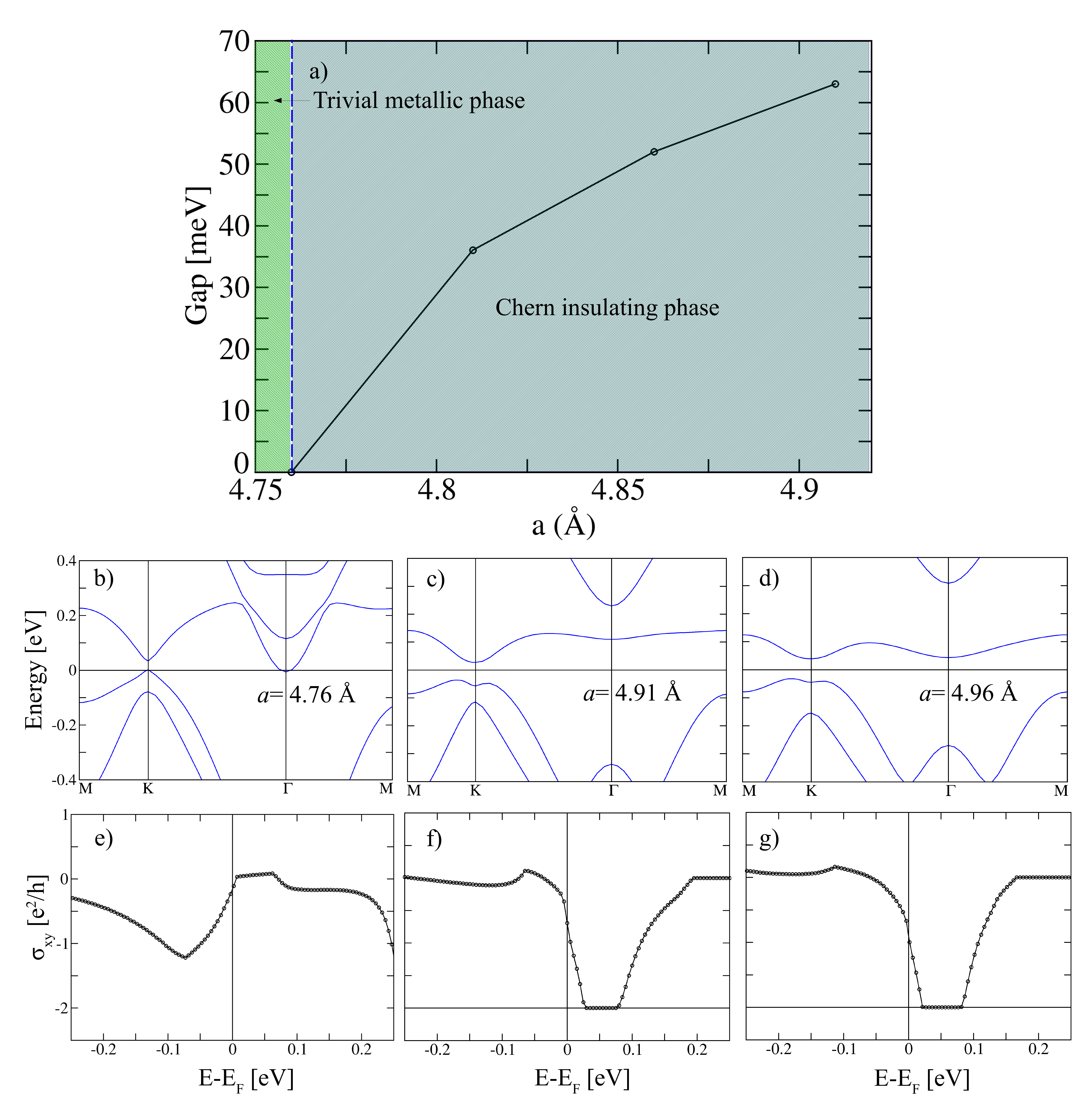}
\caption{\label{fig:Tc-strain} a) Evolution of the band gap for ($Tc_2$O$_3$)$_1$/($Al_2$O$_3$)$_5$(0001) for  $U$\,=\,1.0 eV as a function of the lateral lattice parameter. While for compressive strain the system is trivial metallic, the Chern insulating phase is stabilized  with increasing $a$. In b-d) the band structures and 
e-g) the corresponding anomalous Hall conductivities $\sigma_{xy}$ in units of $e^{2}/h$ as a function of the chemical potential are shown at different $a$ values.}
\end{figure}

A further trend as a function of $U$ is the shift of the gap from K at $U=1.0$ eV to M at $U=3.0$ eV (cf. Fig. \ref{fig:Tc_BC_AHC}c, l). For $U=3.0$ eV the gap closes and the system reverts to a metallic state. A closer examination of the Berry curvature at $U$\,=\,1.0 eV (Fig. \ref{fig:Tc_BC_AHC}f) indicates that the largest contributions to $\Omega(k)$\cite{Wang_Vanderbilt} stem from the SOC-driven band disentanglement of the two occupied majority bands at K as well as the small splitting between the occupied band 152 and unoccupied band 153 (cf. Fig. \ref{fig:Tc_BC_AHC}c) which result in a small denominator. These bands are responsible for the emergence of a broad plateau of $\sigma_{xy}$ just above \ef\ and a QAHI phase with $C=-2$. The predominance of negative contributions to the Berry curvature in Fig. \ref{fig:Tc_BC_AHC}f is in agreement with the negative Chern number. In particular, the largest contribution for $U=2.0$ eV arises along K-M as apparent from the top view of $\Omega(k)$ displayed in Fig. \ref{fig:BC_Tc}a. The topological phase remains stable up to $U$\,=\,2.0 eV and eventually reverts to positive values $\sim e^2/h$ at $U=2.5-3.0$ eV with a stronger contribution of the occupied band 152 (cf. Fig. \ref{fig:Tc_BC_AHC}r). Sharp peaks emerge near the K and M points in $\Omega(k)$ (cf. Fig. \ref{fig:Tc_BC_AHC}o), also visible in the top view of the Berry curvature in Fig. \ref{fig:BC_Tc}b. Finally, at $U$\,=\,3.0 eV (Fig. \ref{fig:Tc_BC_AHC}l) owing to an electron pocket at K and a large hole pocket at $\Gamma$ accompanied by an upward shift of the formerly occupied band, a metallic system is obtained. As a consequence, the large negative contribution to the Berry curvature at K (cf. Fig. \ref{fig:Tc_BC_AHC}f) vanishes completely and drops almost down to zero (Fig. \ref{fig:Tc_BC_AHC}o).  
The dramatic changes of the Berry curvature with $U$ eventually lead to the loss of the nontrivial topological character. The results demonstrate the strong sensitivity of topological quantities on the Coulomb repulsion parameter. A similar transition from a Chern to a trivial Mott insulator was recently predicted for BFPO \cite{Kwan2016}.

We have explored the effect of strain on the CI state: the evolution of band structure and AHC as a function of lateral lattice constant, shown in Fig. \ref{fig:Tc-strain}, demonstrates that while compressive strain closes the gap, the CI phase is stabilized with a broader plateau and enhanced  gap with increasing $a$. 

\subsection{Spin textures}
\begin{figure} [htp!]
\centering                                                          
\includegraphics[width=13.65cm,keepaspectratio]{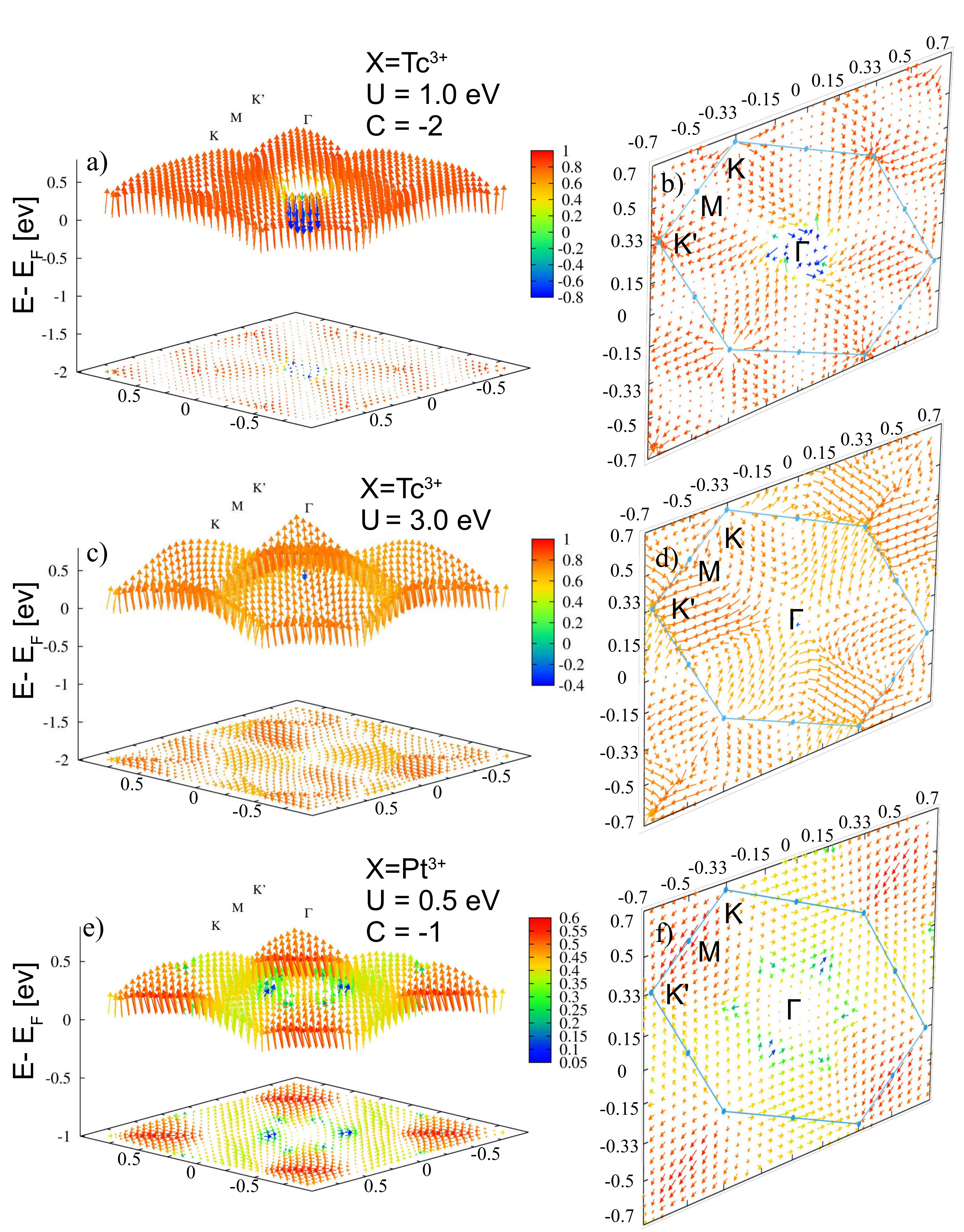}
\caption{Side and top view of the spin textures in $k$-space from GGA\,+\,\textit{U}\,+\,SOC calculations with magnetization along [0001] of band 153  for $X$\,=\,Tc (compare Fig. \ref{fig:Tc_BC_AHC}c)  for a-b) $U$\,=\,1.0 eV and c-d) $U$\,=\,3.0 eV.  e-f) spin textures of band 159 for $X$\,=\,Pt (see \ref{fig:Pd_Pt_SOC_AHC}b). The color scale indicates the projection on the $\hat{z}$-axis with red (blue) indicating parallel (antiparallel) orientation. The top views display the in-plane variation of the spins.}
\label{fig:Spin_texture_Tc_Pt}   
\end{figure}

Since the Berry curvature can be regarded as an effective magnetic field \cite{Weng2015}, it is instructive to relate $\Omega(k)$ to the corresponding spin-textures of the relevant bands, i.e. number 153 for Tc (see Fig. \ref{fig:Tc_BC_AHC}c) and number 159 for Pt (see Fig. \ref{fig:Pd_Pt_SOC_AHC}b). 
%
%

A closer inspection of the spin textures in Fig. \ref{fig:Spin_texture_Tc_Pt}a-b for $U=1.0$ eV reveals an orientation reversal of the out-of-plane spin component $s_z$ from positive in the major part of the BZ to negative around the $\Gamma$ point. The negative polarization around $\Gamma$ is followed by a region of predominantly in-plane components that form a vortex and subsequently an outer region  with positive $s_z$. 
We note that the size of the negatively polarized region and the subsequent formation of a vortex is significantly reduced/almost quenched at $U=3.0$ eV, as displayed in  Fig. \ref{fig:Spin_texture_Tc_Pt}c-d. Interestingly, the spin-texture of the topological phase of $X$\,=\,Pt in Fig. \ref{fig:Spin_texture_Tc_Pt}e-f exhibits only positive $s_z$ values, nearly zero around $\Gamma$ with an in-plane component further away from the BZ center. 

\subsection{Large SOC effect in $Os_{2}$O$_{3}$}
\begin{figure} [hp!]
\hspace*{-1.5cm} 
\includegraphics[width=16.0cm,keepaspectratio]{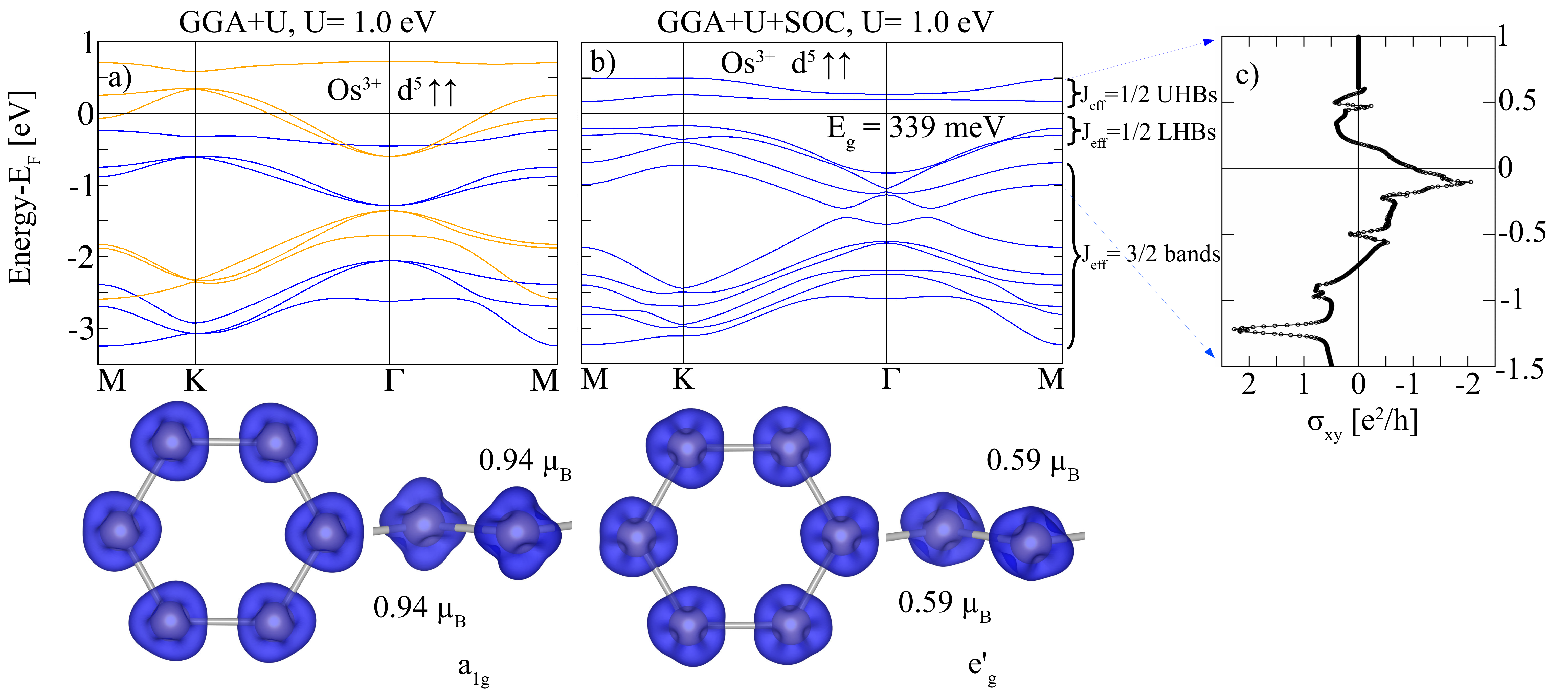}
\caption{Band structures of FM $X$\,=\,Os within  GGA\,+\,\textit{U} with $U$\,=\,1.0 eV a) without and b) with SOC with magnetization along the [0001]-direction.  Note the effect of SOC to split the $5d$ bands in eight (four for each of the two  Os sites in the structure) $j_{\rm eff}=3/2$ and four $j_{\rm eff}=1/2$ bands, with an occupied/unoccupied pair of bands defining the gap. The corresponding spin density distributions are integrated in the energy range of --8 eV to \ef. Same color coding is used as in Fig. \ref{fig:Pd_Pt}. c) anomalous Hall conductivity as a function of the chemical potential. }
\label{fig:Os_SOC}   
\end{figure}
$X$\,=\,Os represents a further example of a large effect of SOC in a $5d$ system. Already within GGA, SOC opens a gap (not shown here) but both spin and orbital moments are quenched and the system is non-magnetic. We note that the topological features of the non-magnetic solution were recently analyzed by Afonso and Pardo \cite{Pardo2015}. Applying even small Hubbard $U=1.0$ eV stabilizes the FM state by $\sim20$ meV per u.c.. The obtained band structures using GGA\,+\,\textit{U} without and with SOC are displayed in Fig. \ref{fig:Os_SOC}a and b. Os ($d^5$) is in the low-spin state with a single hole in the minority spin channel and a magnetic moment of 0.94 \mub\ without SOC. The band structure indicates metallic behavior  with two minority bands crossing \ef. SOC splits the latter and opens a gap of 0.34 eV. In particular the conduction band has a significantly reduced  dispersion. Moreover, the spin moment is reduced to 0.59 \mub, accompanied by a significant orbital moment of 0.34 \mub.

\begin{figure} [ht!]
\centering
\includegraphics[width=6.5cm,keepaspectratio]{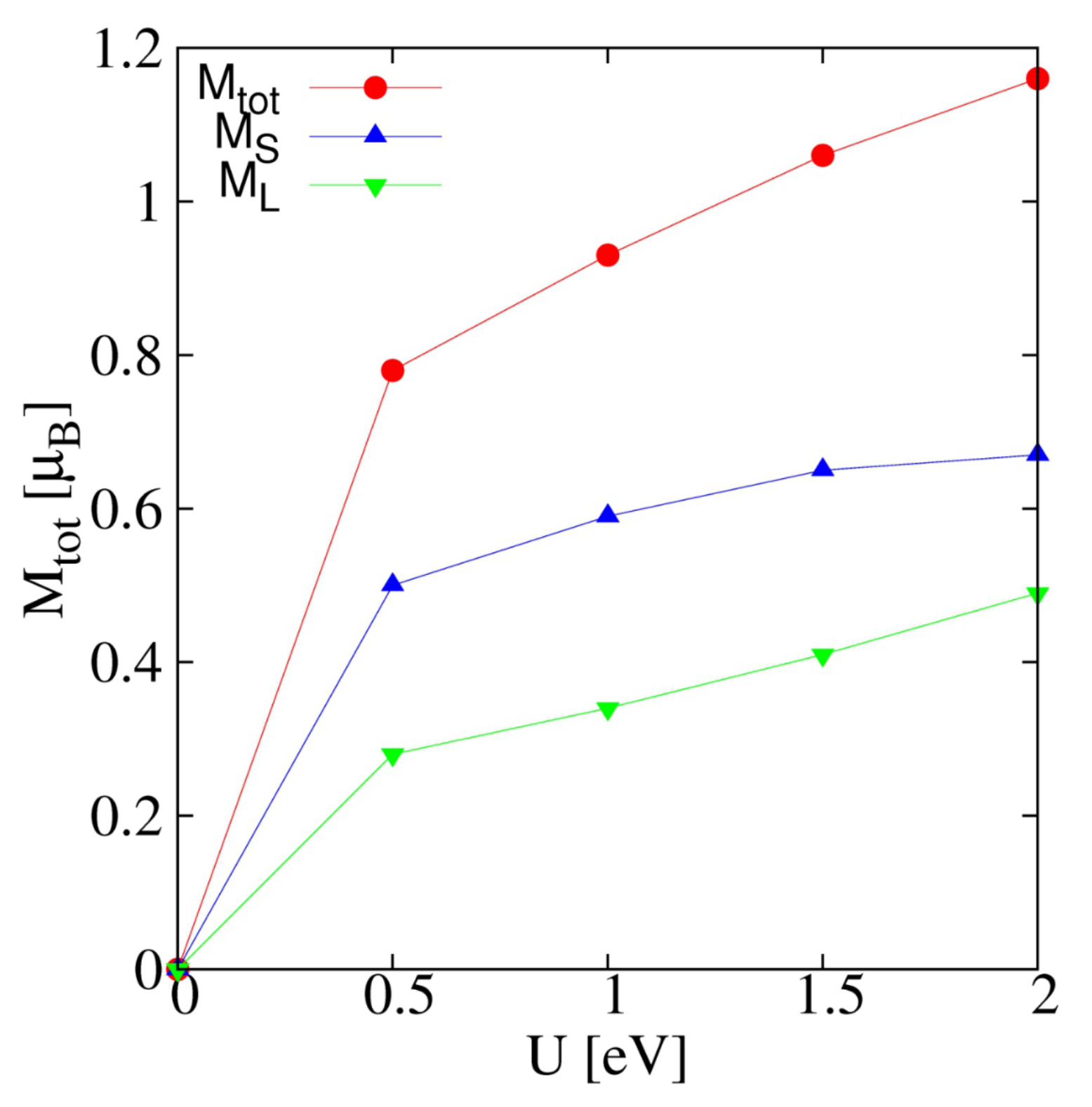}
\caption{Calculated values for spin, orbital moments as well as total moments vs. different values of $U$. A higher value of $U$ leads to an increase of both spin and orbital moments. Since there is no Os orbital moment with antiparallel alignment to the spin moment, a simultaneous increase of the total moment is obtained.}
\label{fig:MS_ML_Os}   
\end{figure}

The evolution of spin and orbital moment as a function of $U$ is shown in Fig. \ref{fig:MS_ML_Os}: After an initial jump between $U=0.0$ (where the spin and orbital moments are quenched) and $U=0.5$ eV with finite values, both exhibit a more moderate increase with further enhancement of $U$. Unlike the LaOsO$_3$ honeycomb bilayer, which was predicted to be a Chern insulator under strain \cite{HongliNQM}, for Os$_2$O$_3$ the Hall conductivity shows large but not-quantized values caused by non-trivial bands, but no plateau at \ef. In the $d^5$ configuration, here with a single hole in the $t_{2g}$ manifold, both the spin and orbital moment are positive, in contrast to other Os-based compounds with $d^1$ configuration like the double perovskite Ba$_2$NaOsO$_6$ \cite{Warren2007}  and the antiferromagnetic $d^1$ system KOsO$_4$ \cite{Warren2014} where the spin moment is almost compensated by the orbital moment. SOC splits the $t_{2g}$ bands (six per Os site, twelve in total) into four filled $j_{\rm eff}=3/2$ and a half-filled $j_{\rm eff}=1/2$ subset. The moderate Hubbard $U$ value opens up a Mott gap separating the upper and lower $j_{\rm eff}$=1/2 Hubbard band (cf. Fig. \ref{fig:Os_SOC}b).  The $j_{\rm eff}$ splitting is similar   to the one of Sr$_2$IrO$_4$ \cite{Kim2008} and occurs due to the large $\zeta_{so}$$\sim0.22$ eV visible in Fig. \ref{fig:Os_SOC}b. Due to Hund's rule coupling the energy is lowered if $\vec{L}$ and $\vec{S}$ are aligned parallel whereas a $d^1$ configuration \cite{Warren2007} yields antiparallel orientation. In 5$d^{5}$ iridates \cite{Arita2012} or for the present compound $X$\,=\,Os, there is one hole in the \tg\ manifold giving rise to an orbital moment $l_{\rm eff}$=1 due to three nearly degenerate orbital configurations. In contrast, the effect of spin-orbit coupling in a 5$d^{3}$ system as in the pyrochlore osmate Cd$_2$Os$_2$O$_7$ \cite{Calder2016} and NaOsO$_3$ \cite{Calder2012} is weak and the orbital moment with $l_{\rm eff}$=0 is quenched because only one orbital configuration is possible for the three \tg\ electrons.

\section{Conclusion}

In conclusion we have explored the interplay between electronic correlation and SOC in $4d$ and $5d$ \xoalo\ SLs with a honeycomb pattern by means of DFT\,+\,\textit{U}\,+\,SOC calculations. The ferromagnetic phases of Tc$_2$O$_3$ and Pt$_2$O$_3$ are identified as Chern insulators with $C$\,=\,--2 and --1 for realistic strengths of Coulomb repulsion and the lateral lattice constants of $a= 4.81$ \AA\ and $a= 4.76$ \AA, respectively. For ($Tc_{2}$O$_{3}$)$_{1}$/(Al$_{2}$O$_{3}$)$_{5}$(0001) with a $d^{4}$ filling a transition from a CI to a trivial metallic state occurs beyond a critical strength $U_c =2.5$ eV. Strain turns out to have a similar effect:  compressive strain renders a metallic trivial state, while the CI phase is stabilized further with tensile strain. 
Moreover, the evolution of Berry curvature and anomalous Hall conductivity as a function of  $U$ is related to the spin texture. In particular, in the CI state ($U =1.0-2.0$ eV) a spin-vortex emerges around $\Gamma$ that is almost quenched for $U=3.0$ eV. In contrast to Pt, the isovalent case of Pd does not host a non-trivial topological phase demonstrating the importance of the interplay and  favorable balance between SOC strength and correlation effects. Last but not least we identified a further case of large SOC effect for ferromagnetic ($Os_{2}$O$_{3}$)$_{1}$/(Al$_{2}$O$_{3}$)$_{5}$(0001) accompanied by the formation of a high orbital moment and a SOC-driven metal-to-insulator transition (MIT) associated with the correlation-induced splitting of the $j_{\rm eff}$=1/2 subset as in the Mott insulator Sr$_2$IrO$_4$ \cite{Kim2008}. 

The growth of corundum thin films on \aalo(0001) has been reported in several experimental studies exploiting different techniques such as molecular beam epitaxy \cite{Chamberlin2014,Kaspar2014,Dennenwaldt2015}, pulsed laser deposition \cite{Popova2008}, helicon plasma-\cite{Takada2008} or r.f. magnetron sputtering~\cite{Gao2017}. Compared to perovskite heterostructures, corundum heterostructures or films contain fewer elements. Since the growth of corundum-based superlattices is feasible, we hope that the theoretical predictions will encourage new experimental efforts for their synthesis. 

\section{Acknowledgement}
We acknowledge discussions with Warren E. Pickett on related systems. We also would like to thank Francisco Mu\~noz and Quansheng Wu for useful discussions on the calculations of spin textures and edge states. We gratefully acknowledge funding by the German Science Foundation within CRC/TRR80, project G3 and computational time at the Leibniz Rechenzentrum, project pr87ro.

\bibliographystyle{model1-num-names}
\bibliography{sample.bib}

\begin{thebibliography}{00}

\bibitem{Weng2015} H. Weng, R. Yu, X. Hu, X. Dai and Z. Fang, Adv. Phys. \textbf{64}, 227 (2015). 
\bibitem{Ren2016} Y. Ren, Z. Qiao, and Q. Niu, Rep. Prog. Phys. \textbf{79}, 066501 (2016).
\bibitem{Haldane} F. D . M. Haldane,  Phys. Rev. Lett. \textbf{61}, 2015 (1988).
\bibitem{Liu_Zhang} C.-X. Liu, X.-L. Qi, X. Dai, Z. Fang and S.-C. Zhang,  Phys. Rev. Lett. \textbf{101}, 146802 (2008).
\bibitem{Yu_Zhang} R. Yu, W. Zhang, H.-J. Zhang, S.-C. Zhang, X. Dai and Z. Fang, 
Science \textbf{329}, 61 (2010).
\bibitem{Fang_Bernevig} C. Fang, M. J. Gilbert and B. A. Bernevig,  Phys. Rev. Lett. \textbf{112}, 046801 (2014).
\bibitem{Zhang2012} H. Zhang, C. Lazo, S. Bl\"ugel, S. Heinze and Y. Mokrousov,  Phys. Rev. Lett. \textbf{108}, 056802 (2012).
\bibitem{Zhou_Liu} M. Zhou, Z. Liu, W. Ming, Z. Wang and F. Liu, Phys. Rev. Lett. \textbf{113}, 236802 (2014). 
\bibitem{Sheng2017} Xian-Lei Sheng and B. K. Nikolic, Phys. Rev. B \textbf{95}, 201402(R) (2017).
\bibitem{Zhang2014} H. Zhang, J.  Wang, G. Xu, Y. Xu and S.-C. Zhang, Phys. Rev. Lett. {\bf 112}, 096804 (2014). 
\bibitem{Garrity2014} K. F. Garrity and D. Vanderbilt, Phys. Rev. B {\bf 90}, 121103(R) (2014).
\bibitem{Huang_Vanderbilt} H. Huang, Z. Liu, H. Zhang, W. Duan and D. Vanderbilt,  Phys. Rev. B \textbf{92}, 161115(R) (2015).
\bibitem{Cai_Gong} T. Cai, X. Li, F. Wang, S. Ju, J. Feng and C.-D. Gong, Nano Lett. \textbf{15}, 6434 (2015).
\bibitem{Lado2016} J. L. Lado, D. Guterding, P. Barone, R. Valenti, and V. Pardo, Phys. Rev. B \textbf{94}, 235111 (2016).
\bibitem{Fiete2015} G.A. Fiete and A. R\"uegg, J. Appl. Phys. \textbf{117}, 172602 (2015).
\bibitem{Xiao2011} D. Xiao, W. Zhu,	Y. Ran,	N. Nagaosa and S. Okamoto, Nature Commun. {\bf 2}, 596 (2011).
\bibitem{Lado2013} J. L. Lado, V. Pardo, and D. Baldomir, Phys. Rev. B \textbf{88}, 155119 (2013).
\bibitem{Okamoto2014} S. Okamoto, W. Zhu, Y. Nomura, R. Arita, D. Xiao, and N. Nagaosa, Phys. Rev. B. \textbf{89}, 195121 (2014).
\bibitem{Doennig2016} D. Doennig, S. Baidya, W. E. Pickett and R. Pentcheva,  Phys. Rev. B. \textbf{93}, 165145 (2016).
\bibitem{caviglia12} A. D. Caviglia, R. Scherwitzl, P. Popovich, W. Hu, H. Bromberger, R. Singla, M. Mitrano, M. C. Hoffmann, S. Kaiser, P. Zubko, S. Gariglio, J.-M. Triscone, M. F\"orst and A. Cavalleri, Phys. Rev. Lett. {\bf 108}, 136801 (2012).
\bibitem{HongliNQM} H. Guo, S. Gangopadhyay, O. K\"oksal, R. Pentcheva and W. E. Pickett, npj Quantum Materials \textbf{2}, 4 (2017).
\bibitem{OKRP2016} O. K\"oksal, S. Baidya, and R. Pentcheva, Phys. Rev. B \textbf{97}, 035126  (2018). 
\bibitem{Kee2017} H. S. Kim and H.-Y. Kee, npj Quantum Mater. \textbf{2}, 20 (2017).
\bibitem{Warren2015} Y.-J. Song, K.-W. Lee, and W. E. Pickett, Phys. Rev. B \textbf{92}, 125109 (2015).
\bibitem{Pardo2015} J. F. Afonso, and V. Pardo, Phys. Rev. B \textbf{92}, 235102 (2015).
\bibitem{VASP} G. Kresse and J. Furthm\"uller, Phys. Rev. B \textbf{54}, 11169 (1996).
\bibitem{PAW} G. Kresse and D. Joubert, Phys. Rev. B. \textbf{59}, 1758 (1999).
\bibitem{GGA_PBE} J. P. Perdew, K. Burke, and M. Ernzerhof, Phys. Rev. Lett. \textbf{77}, 3865 (1996).
\bibitem{Liechtenstein} A.I. Liechtenstein, V.I. Anisimov and J. Zaane, Phys. Rev. B \textbf{52}, R5467 (1995).
\bibitem{Newnham1962} R. E. Newnham and Y. M. de Haan, Zeitschrift f\"ur Kristallographie \textbf{117}, 235-237 (1962).
\bibitem{wien2k} K. Schwarz and P. Blaha, Comput. Mater. Sci. \textbf{28}, 259 (2003).
\bibitem{wannier90} A. A. Mostofi, J. R. Yates, Y.-S. Lee, I. Souza, D. Vanderbilt and N. Marzari,  Comput. Phys. Commun. \textbf{178}, 685 (2008).
\bibitem{romero} A.~Blanca-Romero and R.~Pentcheva, Phys. Rev. B {\bf 84}, 195450 (2011).
\bibitem{Freeland:11} J.~W. Freeland, J.~Liu, M.~Kareev, B.~Gray, J.~W. Kim, P.~Ryan, R.~Pentcheva, and J.~Chakhalian,
Europhys. Lett. {\bf 96}, 57004 (2011).
\bibitem{Doennig2014} D. Doennig, W. E. Pickett and R. Pentcheva, Phys. Rev. B {\bf 89}, 121110 (2014).
\bibitem{Rodriguez2011} E. E. Rodriguez, F. Poineau, A. Llobet, B. J. Kennedy, M. Avdeev, G. J. Thorogood, M. L. Carter, R. Seshadri, D. J. Singh, and A. K. Cheetham, Phys. Rev. Lett.  \textbf{106}, 067201 (2011).
\bibitem{Maxim2011} M. Avdeev, G. J. Thorogood, M. L. Carter, B. J. Kennedy, J. Ting, D. J. Singh, and K. S. Wallwork, J. Am. Chem. Soc. \textbf{133} (6), pp 1654–1657 (2011).
\bibitem{Franchini2011} C. Franchini, T. Archer, Jiangjang He, Xing-Qiu Chen, A. Filipetti, and S. Sanvito, Phys. Rev. B \textbf{83}, 220402(R) (2011).
\bibitem{Georges2012} J. Mravlje, M. Aichhorn, and A. Georges, Phys. Rev. Lett.  \textbf{108}, 197202 (2012).
\bibitem{Vladislav2015} V. Borisov, S. Ostanin, and I. Mertig, Phys.Chem.Chem.Phys. \textbf{17}, 12812 (2015).
\bibitem{Wang_Vanderbilt} X. Wang, J. R. Yates, I. Souza and D. Vanderbilt, Phys. Rev. B \textbf{74}, 195118 (2006).
\bibitem{Warren2007} K.-W. Lee, and W. E. Pickett, Europhys. Lett \textbf{80}, 37008 (2007).
\bibitem{Warren2014} Y.-J. Song, K.-H. Ahn, K.-W. Lee, and W. E. Pickett, Phys. Rev. B \textbf{90}, 245117 (2014).
\bibitem{Kwan2016} Y.-J. Song, K.-H. Ahn, W. E. Pickett, and K.-W. Lee, Phys. Rev. B \textbf{94}, 125134 (2016).
\bibitem{Calder2016} S. Calder, J.G. Vale, N.A. Bogdanov, X. Liu, C. Donnerer, M.H. Upton, D. Casa, A.H. Said, M.D. Lumsden, Z. Zhao, J.-Q. Yan, D. Mandrus, S. Nishimoto, J. van den Brink, J.P. Hill, D.F. McMorrow and A.D. Christianson, Nature Commun. {\bf 7}, 11651 (2016).
\bibitem{Calder2012} S. Calder, V.O. Garlea, D.F. McMorrow, M. D. Lumsden, M.B. Stone, J.C. Lang, J.-W. Kim, J.A. Schlueter, Y.G. Shi, K. Yamaura, Y.S. Sun, Y. Tsujimoto, and A.D. Christianson, Phys. Rev. Lett. \textbf{108}, 257209 (2012).
\bibitem{Kim2008} B.J. Kim, H. Jin, S.J. Moon, J.-Y. Kim, B.-G. Park, C.S. Leem, J. Yu, T.W. Noh, C. Kim, S.-J. Oh, J.-H. Park, V. Durairaj, G. Cao, and E. Rotenberg, Phys. Rev. Lett.  \textbf{101}, 076402 (2008).
\bibitem{Arita2012} R. Arita, J. Kune$\check{s}$, A.V. Kozhevnikov, A.G. Eguiluz, and M. Imada, Phys. Rev. Lett. \textbf{108}, 086403 (2012).
\bibitem{Chamberlin2014} S. E.~Chamberlin, T. C. Kaspar, M. E. Bowden, V. Shutthanandan, B. Kabius, S. Heald, D. J. Keavney, and S. A. Chambers, J. Appl. Phys. {\bf 116}, 233702 (2014).
\bibitem{Kaspar2014} T. C.~Kaspar, S. E. Chamberlin, M. E. Bowden, R. Colby, V. Shutthanandan, S. Manandhar, Y. Wang, P. V. Sushko, and S. A. Chambers, J. Phys.: Cond. Mat. {\bf 26}, 135005 (2014).
\bibitem{Dennenwaldt2015} T.~Dennenwaldt, M. Lübbe, M. Winklhofer, A. M\"uller, M. D\"oblinger, H. Sadat Nabi, M. Gandman, T. Cohen-Hyams, W. D. Kaplan, W. Moritz, R. Pentcheva, C. Scheu, J. Mat. Sci. {\bf 50}, 122 (2015).
\bibitem{Popova2008} E.~Popova, B. Warot-Fonrose, H. Ndilimabaka, M. Bibes, N. Keller, B. Berini, K. Bouzehouane, and Y. Dumont, J. Appl. Phys. {\bf 103}, 093909 (2008).
\bibitem{Takada2008} Y.~Takada, M. Nakanishi, T. Fujiia, and J. Takada, Appl. Phys. Lett. \textbf{92}, 252102 (2008).
\bibitem{Gao2017} Y.~Gao, H. Leiste, M. Stueber, S. Ulrich, J. Cryst. Growth \textbf{457}, 158 (2017).
\bibitem{Suppl} See Supplemental Material at http://link.aps.org/supplemental/... for additional information on topological and structural properties. 
\end{thebibliography}

\end{document}